# *In Vitro* Vascularized Tumor Platform for Modeling Tumor-Vasculature Interactions of Inflammatory Breast Cancer[‡]

Manasa Gadde[1], Caleb Phillips[2*], Neda Ghousifam[3*], Anna G. Sorace[4,5,6], Enoch Wong[1], Savitri Krishnamurthy[7], Anum Syed[1], Omar Rahal[8,9], Thomas E. Yankeelov[1,2,10,11,12], Wendy A. Woodward[8,9+], Marissa Nichole Rylander[1,2,3+]

[1]Department of Biomedical Engineering
The University of Texas at Austin, TX, USA
[2]Oden Institute for Computational and Engineering Sciences
The University of Texas at Austin, TX, USA
[3]Department of Mechanical Engineering
The University of Texas at Austin, TX, USA
[4]Department of Radiology
The University of Alabama at Birmingham, AL, USA
[5]Department of Biomedical Engineering
The University of Alabama at Birmingham, AL, USA
[6]O'Neal Comprehensive Cancer Center
The University of Alabama at Birmingham, AL, USA
[7]Department of Pathology
The University of Texas M. D. Anderson Cancer Center, TX, USA
[8]M. D. Anderson Morgan Welch Inflammatory Breast Cancer Research Program and Clinic
The University of Texas MD Anderson Cancer Center, TX, USA
[9]Department of Experimental Radiation Oncology
The University of Texas M. D. Anderson Cancer Center, TX, USA
[10]Departments of Diagnostic Medicine
The University of Texas at Austin, TX, USA
[11]Department of Oncology
The University of Texas at Austin, TX, USA
[12]Livestrong Cancer Institutes
The University of Texas at Austin, TX, USA
*Co-second authors
+Co-last authors

Manasa Gadde, corresponding author
mgadde@utexas.edu
Caleb Phillips
calebphillips@utexas.edu
Neda Ghousifam
neda.ghousifam@utexas.edu
Dr. Anna G. Sorace
asorace@uabmc.edu
Enoch Wong

---

[‡] Preprint version of this manuscript is available on arXiv.org

enoch.wong@utexas.edu
Savitri Krishnamurthy
skrishna@mdanderson.org
Anum Syed
anum.syed@utexas.edu
Omar Rahal
ORahal@mdanderson.org
Dr. Thomas E. Yankeelov
thomas.yankeelov@utexas.edu
Dr. Wendy A. Woodward
wwoodward@mdanderson.org
Dr. Marissa Nichole Rylander
mnr@austin.utexas.edu

**Acknowledgements**
TEY is a CPRIT Scholar in Cancer Research.

**Funding**
We thank the National Cancer Institute for funding through R01CA186193 and U01CA174706, National Institute of Health for funding through 1R21CA158454-01A1 and R21EB019646, Cancer Prevention Research Institute of Texas Grant RR160005 and the American Cancer Society for funding through RSG-18-006-01-CCE.

**Abstract**

Inflammatory breast cancer (IBC), a rare form of breast cancer associated with increased angiogenesis and metastasis, is largely driven by tumor-stromal interactions with the vasculature and the extracellular matrix (ECM). However, there is currently a lack of understanding of the role these interactions play in initiation and progression of the disease. In this study, we developed the first three-dimensional, *in vitro,* vascularized, breast tumor platform to quantify the spatial and temporal dynamics of tumor-vasculature and tumor-ECM interactions specific to IBC. Platforms consisting of collagen type 1 ECM with an endothelialized blood vessel were cultured with IBC cells, MDA-IBC3 (HER2+) or SUM149 (triple negative), and for comparison to non-IBC cells, MDA-MB-231 (triple negative). An acellular collagen platform with an endothelial blood vessel served as control. SUM149 and MDA-MB-231 platforms exhibited a significantly ($p<0.05$) higher vessel permeability and decreased endothelial coverage of the vessel lumen compared to the control. Both IBC platforms, MDA-IBC3 and SUM149, expressed higher levels of VEGF ($p<0.05$) and increased collagen ECM porosity compared to non-IBC MDA-MB-231 ($p<0.05$) and control ($p<0.01$) platforms. Additionally, unique to the MDA-IBC3 platform, we observed progressive sprouting of the endothelium over time resulting in viable vessels with lumen. The newly sprouted vessels encircled clusters of MDA-IBC3 cells replicating a feature of *in vivo* IBC. The IBC *in vitro* vascularized platforms introduced in this study model well-described *in vivo* and clinical IBC phenotypes and provide an adaptable, high throughout tool for systematically and quantitatively investigating tumor-stromal mechanisms and dynamics of tumor progression.

**Key words:** *In Vitro*, Vasculature, Endothelium, Collagen, Inflammatory Breast Cancer, Triple Negative Breast Cancer, HER2+ Breast Cancer, Microfluidics, Sprouting, Angiogenesis.

# Background

Breast cancer accounts for 15% of newly diagnosed cancer cases in females [1, 2]. Inflammatory breast cancer (IBC) a highly metastatic and aggressive subtype of locally advanced breast cancer and accounts for 10% of all breast cancer related mortality [3-7]. Compared to other metastatic breast cancers, IBC is associated with a median survival of 4 years compared to 10 years in non-inflammatory breast cancer (non-IBC) cases [4]. Approximately 50% of IBC cases lack a tumor mass and present no radiographic evidence. Due to this, the diagnosis of IBC occurs upon clinical manifestation of the disease including pain, redness, and swelling of the breast. At this point, the tumor has advanced to stage III or IV, most patients have lymph node metastases, and 30% of IBC patients exhibit distant metastases compared to 5% for non-IBC [6, 8, 9]. Additionally, contributing to its bleak prognosis, no molecular or histological markers specific to IBC have been identified to distinguish it from other non-IBC breast cancers.

IBC has been shown to be highly angiogenic and metastatic, but a deeper understanding of the diseases dynamics has remained elusive and would enable identification of new diagnostic and therapeutic markers. Current pre-clinical experimental models used to study IBC consist primarily of xenograft animal models, two dimensional (2D) monolayers, and three dimensional (3D) *in vitro* models [10-18]. 2D models do not recapitulate the complex and dynamic nature of the tumor microenvironment which hosts multi-cellular and cell-matrix interactions and evolving biomechanical and chemical features [19-21]. Compared to monolayers, patient derived xenograft (PDX) models are more favorable among researchers as preclinical models for IBC as they provide physiologically relevant tumor microenvironment conditions [7, 22-26]. The Woodward lab has recreated skin invasion and diffuse spread characteristic of IBC in a mouse model with the addition of mesenchymal stem cells [27]. Other examples of xenograft systems for modelling IBC consist of Mary-X and WIBC-9 models. Mary-X established from an IBC patient, recapitulated the human IBC phenotype of extensive lymphovascular invasion of the tumor cell emboli [26], while the WIBC-9 model recreated an invasive ductal carcinoma with a hypervascular structure of solid nests and lymphatic permeation [24]. While PDX models provide a more comprehensive model, determining the influence of specific signaling pathways and microenvironmental stimuli on IBC progression is challenging and frequently cost prohibitive due to the large animal numbers needed. Additionally, dynamic tracking and quantification of tumor presentation and development at a high spatial and temporal resolution is limited in xenograft models. While less common than PDX models, three dimensional (3D) *in vitro* models provide a compromise between 2D and xenograft models as they recapitulate key spatial and physiological facets of the complex tumor microenvironment while maintaining temporal sampling comparable to 2D models. Common 3D *in vitro* IBC models are avascular and consist of culturing IBC monolayers or tumor spheroids on an ECM layer consisting of Matrigel, Culturex, or collagen. [10, 27-35]. These experiments are typically evaluated under static conditions, thereby lacking physiological flow which has been shown to influence tumor response to treatment. [10, 27-29]. There are a number of vascularized 3D breast cancer models that exist ranging from solid gels to spheroids and complex microfluidic models consisting of co-culture of tumor cells with stromal cells such as macrophages, fibroblasts, and pericytes, but these do not incorporate IBC cells and have not been used to study IBC development [36-61]. These platforms have allowed for the exploration of the intricacies of tumor progression and experimental applications in cancer biology, drug delivery, and drug screening. However, due to their inherent limitations including the lack a continuous endothelium, introduction of artificial boundaries and fixtures in the ECM such as pillars for structural stability, they deviate from the *in vivo* tumor architecture [53, 62-65]. Our lab previously established a 3D vascularized microfluidic breast cancer platform incorporated with MDA-MB-231 cells and a vascularized endothelial vessel that addressed the limitation described earlier with existing *in vitro* tumor models. Using this vascularized platform, we determined the relationship between wall shear stress and signaling between cancer and endothelial cells on the vasculature [41, 43, 44] but similar to other *in vitro* tumor models, the platform does not account for the complex tumor dynamics inherent to IBC.

In this study, we described the development and characterization of a versatile, first of its kind, 3D *in vitro* vascularized IBC breast tumor platform as a tool for gaining a deeper understanding of the uniqueness of the IBC phenotype. Specifically, we focused on tumor-vasculature interactions due to the highly angiogenic and metastatic nature of IBC, as well as the significant role that these interactions have in impacting disease phenotype [66-72]. Conditions representative of *in vivo* tumor vasculature interface including physiological flow and associated shear stress were utilized for development of a continuous, aligned, and functional endothelium in the 3D *in vitro* vascularized IBC platform. The vascularized IBC platforms consisted of one of two aggressive IBC cells, MDA-IBC3 (HER2+) and SUM149 (triple negative), and the non-IBC platform consisted of MDA-MB-231 cells (triple negative) cultured within the collagen ECM incorporated with an endothelialized vessel in the center. We quantified the differential effects of the IBC platforms and compared the response, specifically, vascular permeability, endothelial coverage of the vessel lumen, ECM porosity, and cytokine secretion, compared to non-IBC platforms to demonstrate the utility of the platform in the investigation of the spatial and functional interactions not readily quantified in existing *in vivo* IBC models. Additionally, we characterized the spatiotemporal angiogenesis of the MDA-IBC3 platforms and recreated behaviors characteristic of *in vivo* IBC phenotypes including increased angiogenesis, emboli formation, and vascular nesting of tumor emboli. The platforms introduced in this study provide a tool to elucidate unique disease dynamics of IBC and determine the tumor-vasculature interactions driving IBC development and progression.

## Materials and Methods

### Cell Culture

Human breast carcinoma cell line MDA-MB-231(ATCC® HTB-26™) breast carcinoma, human breast inflammatory cancer cells MDA-IBC3 and SUM149, and telomerase-immortalized human microvascular endothelial (TIME) cells were used in this study. MDA-MB-231 and SUM149 are triple negative cell lines while MDA-IBC3 cells are negative for hormone receptors but overexpress human epidermal growth factor receptor 2 (HER2). GFP labeled MDA-MB-231 and mKate labeled TIME cells were a generous gift from Dr. Shay Soker at the Wake Forest Institute for Regenerative Medicine (Winston-Salem, NC). MDA-IBC3 and SUM149 IBC cells labeled with GFP were kindly provided by Dr. Wendy Woodward at MD Anderson Cancer Center (Houston, TX).

MDA-MB-231 cells were cultured in Dulbecco's Modified Eagle's medium, nutrient mixture F-12 (DMEM/F12) (Sigma Aldrich) supplemented with 1% penicillin-streptomycin (P/S) (Invitrogen), and 10 % fetal bovine serum (FBS). MDA-IBC3 and SUM149 cells were cultured in Ham's F-12 media supplemented with 10% FBS, 1% antibiotic-antimycotic, 1 µg/ml hydrocortisone, and 5 µg/ml insulin. TIME cells were cultured in endothelial cell growth medium

2 (PromoCell C-22010). All cell cultures utilized in this study were maintained at 5% $CO_2$ atmosphere and 37°C.

### *In Vitro* 3D Tumor Platform Fabrication

The *in vitro* 3D tumor microfluidic platforms utilized in this study were composed of collagen type I matrix seeded with either GFP labeled MDA-MB-231, MDA-IBC3, or SUM 149 integrated with a hollow channel seeded with mKate labeled TIME cells housed in a polydimethylsiloxane (PDMS) scaffold. Collagen type I extracted from rat tails was prepared following our published protocols to produce stock collagen concentration of 14 mg/ml. Stock collagen was then neutralized with a solution consisting of 10x DMEM, 1N NaOH, and 1x DMEM to yield a final collagen concentration of 7 mg/ml giving comparable stiffness of *in vivo* breast tumors [42, 44, 45, 47, 73]. GFP labeled IBC and non-IBC cells were seeded at a density of $1x10^6$ cells/mL in the 7 mg/ml neutralized collagen solution and polymerized around a 22G needle at 37°C for 25 minutes. After polymerization, the needle was removed, and the resulting hollow void was filled with a solution of $2x10^5$ TIME cells to form an endothelialized vessel lumen. The size of the needle can be varied to mimic vessels of varying sizes in a controllable manner. Flow was introduced using a syringe pump system and a graded flow protocol was used to establish a confluent endothelium as we have previously published [41-44]. Briefly, flow was perfused to expose the endothelium to wall shear stress (WSS) ($\tau$) of 0.01 dyn/cm$^2$ for 36 hours followed by a gradual increase in WSS to 0.1 dyn/cm$^2$ for the following 36 hours and a final increase to 1 dyn/cm$^2$ for 6 hours. Four conditions of the 3D *in vitro* vascularized tumor platforms were created: TIME cell only platform which served as control, and platforms consisting of co-culture of TIME cells with either MDA-MB-231, MDA-IBC3, or SUM149 cells. Endothelial morphology and adhesion, vessel permeability and coverage, matrix porosity, and expression of angiogenic cytokines were characterized in each platform.

### Characterization of Vascularized *In Vitro* IBC and non-IBC Platforms

#### Endothelial Morphology

Endothelial morphology and cell-cell junctions were analyzed by performing immunofluorescent staining for PECAM-1 and F-actin upon completion of the graded flow protocol. PECAM-1 (platelet endothelial cell adhesion molecule-1, green) is expressed at endothelial intercellular junctions and functions in the maintenance of endothelial barrier functions [74]. The staining protocol consisted of perfusing the platforms with 4% paraformaldehyde and 0.5% triton-X 100 for fixation and permeabilization of the cell membranes, respectively. Next, the platforms were incubated in 5% BSA followed by overnight incubation with antibodies for PECAM-1 (Abcam, ab215911) and Rhodamine Phalloidin (Thermo Scientific, R415).

#### Matrix Porosity and Endothelial Adhesion

Scanning electron microscopy (SEM) was performed to determine collagen matrix porosity and observe endothelial adhesion to the collagen matrix. After exposure to the graded flow protocol, the platforms were fixed in an aldehyde mixture overnight at room temperature followed by fixation with osmium on ice for 4 hours. Post fixation, the platforms were dehydrated in an ascending series of ethanol solutions (50-70-95-100%) and then critical point dried by $CO_2$. Platforms were coated with a thin layer of platinum-palladium and SEM imaging was performed with Zeiss Supra40 SEM-Electron Microscope.

### Endothelium Coverage

Vessel volume occupied by TIME cells was quantified using 3D F-actin stained images of the endothelium in each co-culture platform with LASX image processing software. Platforms containing an endothelial vessel but no cancer cells in the surrounding collagen (TIME only platform) served as a control. Reported values for the co-culture platforms were normalized to the control. Significance of the data was verified using one-way ANOVA and a 95% confidence criterion, $p<0.05$.

### Endothelial Permeability

Endothelial vessel permeability as a function of paracrine signaling between tumor and vasculature was determined by perfusing the channels with 70 kDa Oregon green dextran according to our published protocols [41, 75]. After completion of the graded flow protocol for establishing a confluent endothelium, green fluorescent dextran suspended in serum free endothelial growth media (10 µg/ml) was perfused through the platforms with images taken every five minutes. The average fluorescent intensity was measured from the images and used to determine the diffusion permeability coefficient as previously published [41]. Three samples (n=3) were used for each platform condition with the resulting permeability factor expressed as a mean value ± standard deviation. Significance of the data was verified using one-way ANOVA and a 95% confidence criterion.

### Enzyme-linked Immunosorbent Assay

Expression of VEGF, a growth factor known to promote angiogenesis that is excreted from endothelial and tumor cells, was measured using enzyme-linked immunosorbent assays (ELISA) upon completion of the graded flow protocol. 1 ml samples of perfusion media were collected from the flow outlet and ELISA was performed as per manufacturer's protocol (R&D Systems, DVE00). Significance of the data was verified using one-way ANOVA and a 95% confidence criterion.

## Characterization of Angiogenic Sprouting in the *In Vitro* Vascularized MDA-IBC3 Platforms
### Endothelial Sprouting

MDA-IBC3/TIME *in vitro* vascularized platforms were cultured for an additional three weeks following the graded flow protocol in order to characterize endothelial sprouting spatially and temporally. 3D images of the MDA-IBC3/TIME *in vitro* platforms were acquired using Leica TCS SP8 confocal microscope to observe sprout formation and growth. Cross sectional images from the center plane of each channel were used to analyze sprout growth and quantified using ImageJ. Fluorescent intensity histograms for each image were generated using ImageJ's plot profile function. Differences between fluorescence intensity histograms at each time point were quantified using the two-sample Kolmogorov–Smirnov (K-S) statistic, a distance measure between each sample pair's empirical distribution functions. The K-S statistic was calculated between the baseline fluorescence intensity distribution at Day 0 and subsequent imaging time points and significance was determined using $p <0.001$.

### Quantification of Sprout Properties and Vascular Network

Confocal microscopy images acquired on Day 0, 4, 8, 12 and 16 from 3 different platform replicates were analyzed for length of the total vascular network and number of sprouts. Briefly, 11 slices from z-stack of the vessel (~45 µm height) at the center or the widest part of the vessel were analyzed using a Matlab algorithm adapted from work done by Kollmannsberger et al. and Crosby et al. [76-78] to quantify total vascular network and number of sprouts. Analysis of cross sectional areas of the vessel sprouts was performed using methodology developed in house and detailed in supplementary section S1 on Days 0, 4, 8, 12.

### Lumen formation

To confirm the formation of lumen in the newly formed sprouts, platforms were injected with a 20 µl solution of 1 µm Fluoro-Max dyed green aqueous fluorescent microspheres (Thermo Scientific, G0100) on Day 14 and Day 21. The vessels were then imaged using the Leica TCS SP8 confocal microscope and sprouts with beads present were deemed to have formed lumen.

### Cytokine Analyses

Cytokine analyses for CD31, ANG1, ANG2, TGF-α, bFGF, PDGF-bb, EGF, VEGF-A, VEGFR3, VEGF-C, TNF-α, IL-8, IL-6, IL-6 Rα, MMP9, MMP2, MMP13 were performed using a custom human magnetic luminex assay (R&D Systems). Analyses were performed on platform perfusion effluent on Days 0, 7, 14 and 21 according to manufacturer's instructions. Significance of the data was verified using one-way ANOVA and a 95% confidence criterion.

## Results

### *In Vitro* IBC Platform Development

The graded flow preconditioning protocol with a graded increase in WSS from 0.01 dyn/cm$^2$ to 1 dyn/cm$^2$ resulted in a confluent endothelium as shown in Fig. 1a, which shows the evolution of the vascular endothelium in the TIME only *in vitro* vascularized platform. The platforms initiated with a vascular vessel seeded with rounded clusters of TIME cells (0 hour time point) which began to spread out and elongate (24 and 48 hour time points), followed by proliferation and alignment of the cells in the direction of flow to ultimately form the confluent endothelium observed at the 78 hour time point. The resulting endothelium served as the baseline for evaluation of the influence of different cancer cells, IBC and non-IBC, on the surrounding vessel with respect to endothelial morphology, barrier function, and secretion of protumor cytokines (Fig. 1b). In addition to the TIME only *in vitro* vascularized platform, platforms with co-culture of TIME cells with MDA-IBC3, SUM149, and non-IBC MDA-MB-231 breast cancer cells were developed (Fig. 1B). Co-culture of TIME cells with MDA-MB-231 and SUM149 cells resulted in a sparsely covered endothelium evidenced by the presence of large voids in red signal from the endothelium representing areas of the vessel lumen with no endothelial coverage. Both MDA-IBC3/TIME and TIME only *in vitro* vascularized platforms presented a confluent and intact endothelium. The difference in the tumor cells in the platform groups is related to their fluorescent expressions. Emission of the GFP signal from the MDA-IBC3 is much brighter and stronger compared to the other two cells lines. Initial cell seeding shown in Supplementary Fig. A revealed a similar tumor population in the different groups.

## Characterization of *In Vitro* Tumor Platforms

### Endothelial morphology and cell-cell junctions

Endothelial morphology and cell-cell junctions as measured by PECAM-1 and F-actin staining, and SEM are illustrated in Fig. 2a. A compromised endothelium with holes and gaps was observed in the SUM149/TIME and MDA-MB-231/TIME. Staining patterns of PECAM-1 (green, top row) and F-actin (red, middle row) revealed a bright fluorescent signal present continuously across the endothelium in the TIME and MDA-IBC3/TIME *in vitro* vascularized platforms. However, expressions of PECAM-1 and F-actin in SUM149/TIME and MDA-MB-231/TIME were discontinuous with regions of endothelium lacking any signal (pointed out by white arrows) indicating formation of intercellular gaps between neighboring endothelial cells which are typical of a leaky endothelium. Additionally, F-actin staining of MDA-IBC3/TIME platform displayed early signs of angiogenic sprouting with TIME cells starting to bud from the borders of the endothelial vessel (boxed areas) towards MDA-IBC3 cells replicating another important phenomenon characteristic of *in vivo* IBC tumors. High resolution SEM images (bottom row) displayed a tight endothelium with endothelial cell edges overlapping between neighboring cells in the TIME only and the MDA-IBC3/TIME platforms, whereas SUM149/TIME and MDA-MB-231/TIME platforms showed voids between adjacent endothelial cells as denoted by the white arrows.

### Endothelial Lumen Coverage:

Quantitative comparison of endothelial coverage of the lumen, Fig. 2b, exhibited a significant decrease in the endothelium coverage in the SUM149/TIME ($p<0.05$) and MDA-MB-231/TIME ($p<0.01$) platforms, compared to the MDA-IBC3/TIME and control platform as illustrated in Fig. 3. SUM149/TIME had a 1.3 fold and 1.4 fold decrease, and MDA-MB-231/TIME had a 1.5 and 1.6 fold decrease in endothelial coverage compared to control TIME and MDA-IBC3/TIME respectively. There was no significant difference between the TIME and the MDA-IBC3/TIME platforms.

### Endothelial Permeability

Measured effective permeability for TIME, MDA-IBC3/TIME, SUM149/TIME, and MDA-MB-231/TIME platforms were $0.016 \pm 0.002$, $0.019 \pm 0.002$, $0.023 \pm 0.002$, and $0.025 \pm 0.002$ respectively, as portrayed in Fig. 2c. Vascular permeability of the MDA-MB-231/TIME *in vitro* vascularized platforms were statistically significant ($p <0.05$) with 1.6 and 1.3 fold higher permeability than TIME and MDA-IBC3/TIME *in vitro* vascularized platforms respectively. SUM149/TIME *in vitro* vascularized platform also differed significantly from the TIME platforms ($p< 0.05$) with a 1.4 fold increase in permeability. The increased permeability in the MDA-MB-231/TIME and SUM149/TIME platforms confirm the presence of a compromised endothelium and reaffirms the observations from immunofluorescent stained images (Fig. 2a).

### Expression of VEGF and bFGF

ELISA measurements for VEGF and bFGF are illustrated in Fig. 2d with the TIME platform serving as the control. VEGF expression was higher in both the IBC groups (MDA-IBC3, SUM149) compared to non-IBC (MDA-MB-231) and TIME platforms while bFGF was higher in the non-IBC group. VEGF expression was significantly higher ($p < 0.05$) in MDA-IBC3/TIME *in vitro* vascularized platforms compared to the TIME (1.6 times higher) and MDA-MB-231/TIME

(2 times higher) platforms. SUM149/TIME had a higher VEGF expression, 1.5 times, compared to MDA-MB-231/TIME ($p<0.05$). bFGF was expressed highest in the MDA-MB-231/TIME platform, 1.2 times, compared to both the IBC platforms ($p<0.05$).

### Matrix Porosity

Tumor cell morphology and matrix porosity measurements are illustrated in Fig. 3. IBC cells, MDA-IBC3 and SUM149, displayed an epithelial like rounded phenotype while the MDA-MB-231 presented a mesenchymal like phenotype replicating behavior found *in vivo* [79]. Porosity measurements in Fig. 3 revealed significantly more porous collagen ECM in the IBC platforms compared MDA-MB-231/TIME and TIME platforms. SUM149/TIME platforms were 1.5 ($p<0.01$), 1.6 ($p<0.01$), and 1.3 ($p<0.05$) times higher in matrix porosity compared to MDA-MB-231/TIME, TIME only, and MDA-IBC3/TIME *in vitro* vascularized platforms, respectively. MDA-IBC3 *in vitro* platforms also showed an increase in ECM porosity of 1.1 ($p<0.05$) and 1.2 ($p<0.01$) times compared to the MDA-MB-231/TIME and TIME only platforms.

## Reproduction of Relevant IBC Tumor Biology and Phenotypic Comparisons to Published *In Vivo* Models

### Longitudinal Characterization of Vascular Sprouting

Following characterization of the IBC and non-IBC platforms, angiogenic sprouting observed in the MDA-IBC3/TIME was followed for a three week period as illustrated in Fig. 4. This phenomenon was only observed in the presence of MDA-IBC3 cells and not in the presence of SUM149 or MDA-MB-231. Additionally, *in vitro* vascularized platforms composed of BT474, a HER2+ non-IBC cell type, also failed at recreating the extensive angiogenesis present in MDA-IBC3/TIME platforms (data not shown). Fig. 4 reveals the ability of the MDA-IBC3/TIME platforms to promote angiogenic sprouting of the vascular endothelium, formation of MDA-IBC3 tumor emboli, and the capability of the platform for spatiotemporal tracking of the sprouting behavior. Day 0, which represents the endothelium formed after the graded flow protocol, the endothelium exhibited very few sprouts. On Day 4, more sprouts were present with TIME cells extending out from the vessel wall into the collagen. By Days 12 and 16, numerous sprouts formed along the length of the vessel wall with multiple branches and patent lumen (Fig. 4d) invading deeper into the collagen ECM. The sprouts extended towards clusters of MDA-IBC3 and started to encircle these clusters leading to formation and proliferation of MDA-IBC3 emboli as pointed out by the white arrows in the later time points of Day 12 and 16 (Fig. 4a) and in the higher magnification images in Fig. 4c. Vascular encircling of MDA-IBC3 clusters in the *in vitro* platform (Fig. 4f) is reminiscent of IBC tumor behavior *in vivo* in both IBC patients (Fig. 4e) and PDX models of IBC [80, 81]. K-S analysis of vessel sprouting using the center plane of the vessel confirmed a consistent and significant increase in sprout lengths compared to Day 0, $p<0.001$ (Fig. 4b).

### Lumen Formation

Lumen presence in the new angiogenic sprouts were confirmed if green fluorescent microspheres were observed. In TIME only platforms and acellular platforms without an endothelialized vessel (data not shown), perfusion of 1 µm green fluorescent microsphere through the vessel resulted in their aggregation at the vessel walls without entering the surrounding collagen ECM. Fig. 5a and b, confocal images of vessel sprouts taken on Day 14 and 21 reveal the presence of fluorescent microsphere in the vessel sprouts and not in the surrounding collagen matrix indicating the formation of a lumen that allowed for the beads to be transported from the main vessel. By Day 21, we observed an increase in the number of sprouts positive for the presence of the green fluorescent microspheres.

### Quantification of Sprout Properties and Vascular Network

Total length of the vascular network, number of sprouts, and analysis of sprout area along the length of the sprouts are shown in Fig. 6. For determining the number of newly formed sprouts and the total network length, a 45 µm section at the center of the vessel was used (Fig. 6a). The computational recreation of the vascular network from the 45 µm region of interest and the corresponding measurements of number of sprouts and total vascular network are shown in Fig. 6b and c respectively. As expected, the total vascular network and number of sprouts at each subsequent time point increased indicating continuous angiogenic sprouting (Fig. 6b and c). While the growth trends in vascular network and number of newly formed sprouts at each time point are similar between the platforms, the number of sprouts and length of network varies between the different platforms. The analysis for the sprout areas along the sprout lengths showed an increase in the sprout area at later time points. Each sprout was analyzed 100, 200, 300, and 400µm from the edge of the vessel as depicted in the schematic in Fig. 6d. At each distance from the vessel, the number of sprouts of varying area: 100, 200, 300, 400, 500, and 1000 µm$^2$ which correspond to a cross sectional diameter of approximately 11 µm, 16 µm, 20 µm, 23 µm, 25 µm, and 36 µm were counted. On Day 4, the longest vessel was measured 300 µm away from the edge of the vessel. At later time points of Day 8 and Day 12, sprouts were present 400 µm away from the vessel. With time, larger vessels of areas 1000 µm$^2$ which correlate to larger sprouts with lumen were detected and in accordance with observations of lumen formation in image Fig. 6f taken on Day 12 in the *in vitro* MDA-IBC3/TIME platform.

### Cytokine Analyses of Vascular Sprouting

Cytokine analysis of the perfusion media at the outlet was measured at multiple time points illustrated in Fig. 7 and performed to understand the driving factors behind the sustained angiogenic sprouting and kinetics of their expression. VEGF-A, ANG-2, PDGF-bb, IL-8, IL-6, and MMP2 expressions were significantly higher ($p<0.05$) on Day 21 compared to the earlier timepoints. VEGF-A expression was higher at the later timepoints (Day 7, 14, and 21) compared to Day 0 and IL-8 expression increased significantly on Day 14 and 21 compared to Day 0. bFGF and EGF both showed a similar trend with expression peaking on Day 7 ($p<0.05$) and then decreasing back to levels comparable to Day 0 on Days 14 and 21.

## Discussion

In this study, we developed the first 3D *in vitro* vascularized tumor platform to model the unique interactions of IBC cells (SUM149 and MDA-IBC3) with the vasculature and ECM in a dynamic and spatial manner and compared the observations to non-IBC (MDA-MB-231) cells cultured in the platform. Tumor specific *in vivo* responses including increased vascular permeability, ECM remodeling, and vessel sprouting as a result of the tumor-vasculature and tumor-ECM interactions were reproduced and we showed a differential response of the three different cells lines (IBC vs non-IBC and HER2+ vs triple negative) in modulating these behaviors. After identifying the differences between the tumor cells, we investigated the vascular sprouting nature of HER2+ MDA-IBC3 with the platform providing the first opportunity to spatially observe and quantify this behavior *in vitro* and were able to recreate and validate previously published *in vivo* phenotypes including endothelial sprouting, and vascular encircling of tumor emboli.

## Characterization of *In Vitro* Tumor Platforms

For comparison between IBC and non-IBC as well as HER2+ and triple negative IBC cells, we investigated the influence of cell type on vascular permeability of the endothelium, endothelial coverage of the vessel lumen, expression of angiogenic factors VEGF and bFGF as well as remodeling of the collagen ECM. Tumor vasculature is characterized by the presence of leaky blood vessels which has been implicated in inefficient delivery of chemotherapies as well as playing a crucial role in tumor intravasation [82-87]. In this study, we demonstrated that the presence of triple negative, both IBC (SUM149) and non-IBC (MDA-MB-231) cells compromised the vasculature with formation of large pores and gaps in the endothelium, increased vascular permeability, and decreased endothelial coverage of the vessel lumen. Vascular permeability, a measurement of the integrity of the endothelium, was significantly higher in the platforms with the triple negative cancer cells (SUM149 and MDA-MB-231) regardless of IBC or non-IBC status and is in accordance with results from multiple groups where introduction of highly invasive tumor cells increased permeability of the endothelium [53, 88-95]. In addition to vascular permeability, these same platforms exhibited a significant decrease in endothelial population correlating with previous studies that showed direct contact between triple negative MDA-MB-231 and endothelial cells disrupted endothelial monolayers and resulted in anoikis of endothelial cells allowing for dextran to cross into the collagen unhindered correlating with results seen in other experimental studies [94, 96-102]. In contrast to the triple negative cells, HER2+ MDA-IBC3, did not significantly alter the endothelium barrier function and maintained a confluent endothelium. Bright patches of red fluorescent signal in the F-actin stained images, as well as the increased coverage of the endothelium in the MDA-IBC3/TIME *in vitro* vascularized platform, suggest the presence of a larger endothelial population consistent with previous studies that demonstrated a strong association between IBC and increased proliferation levels of endothelial cells [3, 14, 80, 103, 104].

While vascular permeability and endothelial coverage of the lumen were not deterministic factors between IBC and non-IBC cells, expression of the proangiogenic factor VEGF and ECM remodeling were significantly higher in the IBC platforms regardless of receptor status. Both the HER2+ MDA-IBC3 and triple negative SUM149 platforms expressed increased amounts of VEGF and were more active in remodeling the collagen ECM as evidenced by the increased ECM porosity. van Golen et al. determined increased levels of VEGF mRNA in IBC tumors vs non-IBC tumors [105] corresponding with the increased levels of VEGF expression in both the IBC MDA-IBC3 and SUM149 *in vitro* vascularized platforms. Along with being highly angiogenic, IBC tumors are also highly invasive. Analysis of SEM images of the acellular collagen matrix (data not shown) revealed a pore size of ~1 µm, much smaller than cell width. Pore sizes smaller than a cell's width induce cellular degradation of the ECM through secretion of matrix metalloproteinases (MMPs) to allow for motility of cancer cells

[106-111]. Al-Raawi *et al* found an overexpression of MMPs by IBC carcinoma tissues [112] which are involved in degradation of collagen I and widening of pore size to allow for cell migration and invasion [109, 111, 113-116]. Rizwan *et al* demonstrated an increased migratory and invasive behavior in IBC cells as well as increased levels of MMP9 compared to MDA-MB-231 cells [117]. Higher proteolytic activity of IBC breast tumors compared to non-IBC tumors coincides with the increased matrix porosity in the SUM149 and MDA-IBC3 *in vitro* vascularized platforms.

### Reproduction of Relevant IBC Tumor Biology and Phenotypic Comparisons to Published *In Vivo* Models

To the best of our knowledge, this is the first demonstration of the dynamics of vascular sprouting in a 3D *in vitro* platform sustained through interactions between tumor and endothelial cells without the influence of any exogenous supplements or additional stromal cells. IBC is characterized as highly angiogenic with a significantly higher population of tumor infiltrating and proliferating endothelial cells compared to non-IBC cells [80, 104] which is evidenced with the sustained angiogenesis occurring and directed towards tumor cells in our MDA-IBC3 *in vitro* platforms. We performed further studies to confirm whether the vascular angiogenesis seen in the HER2+ MDA-IBC3 was due to HER2+ status or HER2+ and IBC status. MDA-IBC3 platforms were compared to HER2+ non-IBC BT474 vascularized platforms, and observed vascular sprouting only in the MDA-IBC3 platforms (data not shown), revealing the angiogenic behavior attributed to the cells being both HER2+ and IBC. Along with vessel sprouting, we saw the formation and growth of MDA-IBC3 emboli enveloped by newly formed vascular vessels which is characteristic of *in vivo* IBC tumors. Histology samples of IBC tumors and 3D spheroid IBC assays reveal tightly packed clusters of IBC cells similar to MDA-IBC3 emboli developed in the *in vitro* platforms [31, 118]. In an invasion independent metastasis mechanism proposed by Sugino et al., tumor clusters accessed blood vessels by being surrounded by the vessels rather than intravasation, similar to behavior seen in the MDA-IBC3/TIME vascularized *in vitro* breast tumor platforms (Fig. 4C) [119]. Published work by Mahooti et al. describe a phenotype of encircling vasculogenesis in the Mary-X IBC mouse model [81], behavior reproduced by the endothelial sprouts in our *in vitro* platform encircling MDA-IBC3 cells in the matrix demonstrating this *in vivo* phenotype (Fig. 4). Analysis of cytokine expression in the MDA-IBC3 platforms revealed a significant increase in the proangiogenic factors by Day 21 compared to Day 0 associated with significant amount of angiogenesis occurring at the later time point. The highest expression of most of the measured factors, occurred on Day 21 but bFGF and EGF both displayed similar trends in expression levels with the highest expression on Day 7. Additionally, we determined VEGF, an important angiogenic factor [120, 121], to be a key contributor of angiogenesis in our system as continued increase in expression of VEGF paralleled the increase in angiogenic response in the MDA-IBC3 platform. We also confirmed for lumen development in the newly formed sprouts as an indicator of viable vessels and saw the formation of larger and longer vessels over time reminiscent of angiogenic processes. Along with lumen formation, we determined the presence of a larger population of vessels with patent lumen extending further out into the collagen. The trends observed in both the sprout area, number of sprouts and length of the total vascular network showed an increase with each subsequent point yet there were no significant differences between time points. Upon looking at the trends of the individual platforms, we observed one platform presented a much larger vascular network compared to the other two leading to large variation between the platforms.

There are some limitations to our study and the tumor platforms presented. While the *in vitro* platforms developed in this study do not encompass the entire complexity of the tumor microenvironment and utilize immortalized endothelial cells, they recapitulate key IBC characteristics in their current form not available with existing platforms and provide an initial insight into the behavior of aggressive breast tumors and enabling us to recapitulate key phenotypic behaviors specific to IBC. Future experiments utilizing this platform can be expanded to incorporate stromal and immune cells known to influence tumor behavior. Macrophages have been shown to be a key player in driving IBC phenotype and therefore will be an important factor to include in future studies [30, 33, 122]. Other cells for incorporation in the platform include mesenchymal stem cells, adipocytes and fibroblasts all of which have shown to also contribute to IBC phenotype. Additionally, we acknowledge that the size of the endothelial vessels is larger than the size of *in vivo* microvasculature. Platforms can be adapted to present a more comparable vessel with the use of smaller gauge needles for formation of the cylindrical vessels as published in our previous work [44].

## Conclusion

The 3D *in vitro* vascularized IBC platforms presented in this work enabled us to dynamically characterize and model the breast tumor-vascular interactions, as well as determine the spatiotemporal response of these interactions on vascular permeability and matrix porosity not possible with existing *in vitro* or *in vivo* models. The platforms provide a robust and cost-effective means to systematically and quantitatively investigate IBC in a controlled and replicable manner compared to the current standard of using PDX models. Using the *in vitro* platform, we determined IBC cells were more active in remodeling of the collagen ECM as well as secretion of proangiogenic and tumorigenic factor VEGF compared to non-IBC MDA-MB-231, revealing potential targets for IBC therapeutics. For the first time, we induced angiogenic sprouting of the vascular endothelium and vascular surrounding of tumor emboli (unique behaviors of IBC tumors) purely through tumor-endothelial cell interactions and characterized sprouting in a spatial and temporal fashion. Furthermore, our system captures blood vessel leakiness and increased matrix porosity, representative behavior of *in vivo* invasive tumors. Compared to current 3D *in vitro* tumor models that focus on recreating specific stages of tumor progression, our tumor platforms enable the temporal study of various stages in breast cancer progression, including early signs of angiogenesis as well as modulation of tumor ECM and vasculature for migration and metastasis. With the vascularized IBC tumor platforms, behavioral variations that are representative of *in vivo* tumors can be identified and distinguished as a result of the different breast cancer cells. This platform allows for spatiotemporal imaging and identification of biological proteins and responses which may play a direct role in tumorigenesis and vascularization *in vivo*. These platforms represent a useful tool for studying various aggressive breast cancers whose phenotype is driven by tumor-stromal-vascular interactions. These platforms can be further expanded to investigate increasingly complex cell type interactions, thereby providing a tool to further decipher the mechanisms behind development of these tumors.

**List of abbreviations:**
ECM: Extracellular Matrix
IBC: Inflammatory Breast Cancer
2D: 2 Dimensional
3D: 3 Dimensional
TIME: Telomerase Immortalized Microvascular Endothelial
GFP: Green Fluorescent Protein
RFP: Red Fluorescent Protein
WSS: Wall Shear Stress
VEGF: Vascular Endothelial Growth Factor
HER2: Human Epidermal Growth Factor Receptor 2
PECAM-1: Platelet Endothelial Cell Adhesion Molecule-1

# References

1. *Female Breast Cancer - Cancer Stat Facts*.
2. *CDC - Breast Cancer Statistics*. 2017 2017-06-26T16:17:11Z.
3. Costa, R., C.A. Santa-Maria, G. Rossi, B.A. Carneiro, Y.K. Chae, W.J. Gradishar, et al. Developmental therapeutics for inflammatory breast cancer: Biology and translational directions. Oncotarget 8, 12417, 2017.
4. Hance, K.W., W.F. Anderson, S.S. Devesa, H.A. Young, and P.H. Levine. Trends in inflammatory breast carcinoma incidence and survival: the surveillance, epidemiology, and end results program at the National Cancer Institute. J Natl Cancer Inst 97, 966, 2005.
5. Fouad, T.M., A.M.G. Barrera, J.M. Reuben, A. Lucci, W.A. Woodward, M.C. Stauder, et al. Inflammatory breast cancer: a proposed conceptual shift in the UICC-AJCC TNM staging system. Lancet Oncol 18, e228, 2017.
6. Fouad, T.M., T. Kogawa, J.M. Reuben, and N.T. Ueno, *The role of inflammation in inflammatory breast cancer*. *Inflammation and Cancer.* Springer. 2014. 53-73.
7. Lim, B., W.A. Woodward, X. Wang, J.M. Reuben, and N.T. Ueno. Inflammatory breast cancer biology: the tumour microenvironment is key. Nature Reviews Cancer, 1, 2018.
8. Fernandez, S.V., F.M. Robertson, J. Pei, L. Aburto-Chumpitaz, Z. Mu, K. Chu, et al. Inflammatory breast cancer (IBC): clues for targeted therapies. Breast cancer research and treatment 140, 23, 2013.
9. Giordano, S.H. and G.N. Hortobagyi. Clinical progress and the main problems that must be addressed. Breast Cancer Research 5, 284, 2003.
10. Lehman, H.L., E.J. Dashner, M. Lucey, P. Vermeulen, L. Dirix, S.V. Laere, et al. Modeling and characterization of inflammatory breast cancer emboli grown in vitro. International Journal of Cancer 132, 2283, 2013.
11. van Golen, K.L., L. Bao, M.M. DiVito, Z. Wu, G.C. Prendergast, and S.D. Merajver. Reversion of RhoC GTPase-induced inflammatory breast cancer phenotype by treatment with a farnesyl transferase inhibitor. Mol Cancer Ther 1, 575, 2002.
12. van Golen, K.L., L.W. Bao, Q. Pan, F.R. Miller, Z.F. Wu, and S.D. Merajver. Mitogen activated protein kinase pathway is involved in RhoC GTPase induced motility, invasion and angiogenesis in inflammatory breast cancer. Clin Exp Metastasis 19, 301, 2002.
13. van Golen, K.L., Z.F. Wu, X.T. Qiao, L.W. Bao, and S.D. Merajver. RhoC GTPase, a novel transforming oncogene for human mammary epithelial cells that partially recapitulates the inflammatory breast cancer phenotype. Cancer Res 60, 5832, 2000.
14. van Uden, D.J., H.W. van Laarhoven, A.H. Westenberg, J.H. de Wilt, and C.F. Blanken-Peeters. Inflammatory breast cancer: an overview. Crit Rev Oncol Hematol 93, 116, 2015.
15. Charafe-Jauffret, E., C. Ginestier, F. Iovino, C. Tarpin, M. Diebel, B. Esterni, et al. Aldehyde dehydrogenase 1-positive cancer stem cells mediate metastasis and poor clinical outcome in inflammatory breast cancer. Clin Cancer Res 16, 45, 2010.
16. Klopp, A.H., L. Lacerda, A. Gupta, B.G. Debeb, T. Solley, L. Li, et al. Mesenchymal stem cells promote mammosphere formation and decrease E-cadherin in normal and malignant breast cells. PLoS One 5, e12180, 2010.
17. Silvera, D., R. Arju, F. Darvishian, P.H. Levine, L. Zolfaghari, J. Goldberg, et al. Essential role for eIF4GI overexpression in the pathogenesis of inflammatory breast cancer. Nat Cell Biol 11, 903, 2009.
18. Silvera, D. and R.J. Schneider. Inflammatory breast cancer cells are constitutively adapted to hypoxia. Cell Cycle 8, 3091, 2009.

19. Jang, S.H., M.G. Wientjes, D. Lu, and J.L.-S. Au. Drug delivery and transport to solid tumors. Pharmaceutical research 20, 1337, 2003.
20. Kim, B.J. and M. Wu. Microfluidics for mammalian cell chemotaxis. Annals of biomedical engineering 40, 1316, 2012.
21. Trédan, O., C.M. Galmarini, K. Patel, and I.F. Tannock. Drug resistance and the solid tumor microenvironment. Journal of the National Cancer Institute 99, 1441, 2007.
22. Alpaugh, M.L., J.S. Tomlinson, S. Kasraeian, and S.H. Barsky. Cooperative role of E-cadherin and sialyl-Lewis X/A-deficient MUC1 in the passive dissemination of tumor emboli in inflammatory breast carcinoma. Oncogene 21, 3631, 2002.
23. Robertson, F.M., K. Chu, S.V. Fernandez, Z. Mu, X. Zhang, H. Liu, et al. Genomic Profiling of Pre-Clinical Models of Inflammatory Breast Cancer Identifies a Signature of Epithelial Plasticity and Suppression of TGFÃŽÂ² Signaling. Journal of Clinical & Experimental Pathology 2, 1, 2012.
24. Shirakawa, K., H. Kobayashi, J. Sobajima, D. Hashimoto, A. Shimizu, and H. Wakasugi. Inflammatory breast cancer: Vasculogenic mimicry and its hemodynamics of an inflammatory breast cancer xenograft model. Breast Cancer Research 5, 136, 2003.
25. Wurth, R., K. Tarn, D. Jernigan, S.V. Fernandez, M. Cristofanilli, A. Fatatis, et al. A Preclinical Model of Inflammatory Breast Cancer to Study the Involvement of CXCR4 and ACKR3 in the Metastatic Process. Translational Oncology 8, 358, 2015.
26. Alpaugh, M.L., J.S. Tomlinson, Z.M. Shao, and S.H. Barsky. A novel human xenograft model of inflammatory breast cancer. Cancer Research 59, 5079, 1999.
27. Lacerda, L., B.G. Debeb, D. Smith, R. Larson, T. Solley, W. Xu, et al. Mesenchymal stem cells mediate the clinical phenotype of inflammatory breast cancer in a preclinical model. Breast Cancer Research 17, 42, 2015.
28. Lacerda, L., J.P. Reddy, D. Liu, R. Larson, L. Li, H. Masuda, et al. Simvastatin radiosensitizes differentiated and stem-like breast cancer cell lines and is associated with improved local control in inflammatory breast cancer patients treated with postmastectomy radiation. Stem cells translational medicine 3, 849, 2014.
29. Mohamed, M.M., D. Cavallo-Medved, and B.F. Sloane. Human monocytes augment invasiveness and proteolytic activity of inflammatory breast cancer. Biological chemistry 389, 1117, 2008.
30. Allen, S.G., Y.-C. Chen, J.M. Madden, C.L. Fournier, M.A. Altemus, A.B. Hiziroglu, et al. Macrophages Enhance Migration in Inflammatory Breast Cancer Cells via RhoC GTPase Signaling. Scientific Reports 6, 39190, 2016.
31. Arora, J., S.J. Sauer, M. Tarpley, P. Vermeulen, C. Rypens, S.V. Laere, et al. Inflammatory breast cancer tumor emboli express high levels of anti-apoptotic proteins: use of a quantitative high content and high-throughput 3D IBC spheroid assay to identify targeting strategies. Oncotarget 8, 25848, 2017.
32. Hoffmeyer, M.R., K.M. Wall, and S.F. Dharmawardhane. In vitro analysis of the invasive phenotype of SUM 149, an inflammatory breast cancer cell line. Cancer Cell Int 5, 11, 2005.
33. Mohamed, M.M., E.A. El-Ghonaimy, M.A. Nouh, R.J. Schneider, B.F. Sloane, and M. El-Shinawi. Cytokines secreted by macrophages isolated from tumor microenvironment of inflammatory breast cancer patients possess chemotactic properties. The International Journal of Biochemistry & Cell Biology 46, 138, 2014.
34. Morales, J. and M.L. Alpaugh. Gain in cellular organization of inflammatory breast cancer: A 3D in vitro model that mimics the in vivo metastasis. BMC Cancer 9, 462, 2009.
35. Nokes, B.T., H.E. Cunliffe, B. LaFleur, D.W. Mount, R.B. Livingston, B.W. Futscher, et al. In Vitro Assessment of the Inflammatory Breast Cancer Cell Line SUM 149: Discovery of 2 Single Nucleotide Polymorphisms in the RNase L Gene. Journal of Cancer 4, 104, 2013.

36. Huang, C.-J. and Y.-C. Chang. Construction of Cell–Extracellular Matrix Microenvironments by Conjugating ECM Proteins on Supported Lipid Bilayers. Frontiers in Materials 6, 2019.
37. Ma, Y.-H.V., K. Middleton, L. You, and Y. Sun. A review of microfluidic approaches for investigating cancer extravasation during metastasis. Microsystems & Nanoengineering 4, 17104, 2018.
38. Ozcelikkale, A., H.-r. Moon, M. Linnes, and B. Han. In vitro Microfluidic Models of Tumor Microenvironment to Screen Transport of Drugs and Nanoparticles. Wiley interdisciplinary reviews. Nanomedicine and nanobiotechnology 9, 2017.
39. Pouliot, N., H.B. Pearson, and A. Burrows, *Investigating Metastasis Using In Vitro Platforms*. Landes Bioscience. 2013.
40. Rhodes, J.M. and M. Simons. The extracellular matrix and blood vessel formation: not just a scaffold. Journal of Cellular and Molecular Medicine 11, 176, 2007.
41. Buchanan, C.F., S.S. Verbridge, P.P. Vlachos, and M.N. Rylander. Flow shear stress regulates endothelial barrier function and expression of angiogenic factors in a 3D microfluidic tumor vascular model. Cell Adh Migr 8, 517, 2014.
42. Buchanan, C.F., E.E. Voigt, C.S. Szot, J.W. Freeman, P.P. Vlachos, and M.N. Rylander. Three-dimensional microfluidic collagen hydrogels for investigating flow-mediated tumor-endothelial signaling and vascular organization. Tissue Eng Part C Methods 20, 64, 2014.
43. Gadde, M., D. Marrinan, R.J. Michna, and M.N. Rylander, *Three Dimensional In Vitro Tumor Platforms for Cancer Discovery*, S. Soker and A. Skardal. *Tumor Organoids.* Springer International Publishing. 2018. 71-94.
44. Michna, R., M. Gadde, A. Ozkan, M. DeWitt, and M. Rylander. Vascularized microfluidic platforms to mimic the tumor microenvironment. Biotechnology and Bioengineering, 2018.
45. Szot, C.S., C.F. Buchanan, J.W. Freeman, and M.N. Rylander. In vitro angiogenesis induced by tumor-endothelial cell co-culture in bilayered, collagen I hydrogel bioengineered tumors. Tissue Eng Part C Methods 19, 864, 2013.
46. Buchanan, C.F., C.S. Szot, T.D. Wilson, S. Akman, L.J. Metheny-Barlow, J.L. Robertson, et al. Cross-talk between endothelial and breast cancer cells regulates reciprocal expression of angiogenic factors in vitro. Journal of Cellular Biochemistry 113, 1142, 2012.
47. Szot, C.S., C.F. Buchanan, J.W. Freeman, and M.N. Rylander. 3D in vitro bioengineered tumors based on collagen I hydrogels. Biomaterials 32, 7905, 2011.
48. Sontheimer-Phelps, A., B.A. Hassell, and D.E. Ingber. Modelling cancer in microfluidic human organs-on-chips. Nature Reviews Cancer 19, 65, 2019.
49. Shang, M., R.H. Soon, C.T. Lim, B.L. Khoo, and J. Han. Microfluidic modelling of the tumor microenvironment for anti-cancer drug development. Lab on a Chip 19, 369, 2019.
50. Ko, J., J. Ahn, S. Kim, Y. Lee, J. Lee, D. Park, et al. Tumor spheroid-on-a-chip: a standardized microfluidic culture platform for investigating tumor angiogenesis. Lab on a Chip 19, 2822, 2019.
51. Osaki, T., J.C. Serrano, and R.D. Kamm. Cooperative Effects of Vascular Angiogenesis and Lymphangiogenesis. Regenerative engineering and translational medicine 4, 120, 2018.
52. Malandrino, A., R.D. Kamm, and E. Moeendarbary. In Vitro Modeling of Mechanics in Cancer Metastasis. ACS Biomaterials Science & Engineering 4, 294, 2018.
53. Tsai, H.F., A. Trubelja, A.Q. Shen, and G. Bao. Tumour-on-a-chip: microfluidic models of tumour morphology, growth and microenvironment. J R Soc Interface 14, 2017.
54. Duinen, V.v., A.v.d. Heuvel, S.J. Trietsch, H.L. Lanz, J.M.v. Gils, A.J.v. Zonneveld, et al. 96 perfusable blood vessels to study vascular permeability in vitro. Scientific Reports 7, 1, 2017.
55. Kim, S., M. Chung, J. Ahn, S. Lee, and N. Li Jeon. Interstitial flow regulates the angiogenic response and phenotype of endothelial cells in a 3D culture model. Lab on a Chip 16, 4189, 2016.

56. Kim, J., M. Chung, S. Kim, D.H. Jo, J.H. Kim, and N.L. Jeon. Engineering of a Biomimetic Pericyte-Covered 3D Microvascular Network. PLOS ONE 10, e0133880, 2015.
57. Jeon, J.S., S. Bersini, M. Gilardi, G. Dubini, J.L. Charest, M. Moretti, et al. Human 3D vascularized organotypic microfluidic assays to study breast cancer cell extravasation. Proceedings of the National Academy of Sciences 112, 214, 2015.
58. Bersini, S., J.S. Jeon, G. Dubini, C. Arrigoni, S. Chung, J.L. Charest, et al. A Microfluidic 3D In Vitro Model for Specificity of Breast Cancer Metastasis to Bone. Biomaterials 35, 2454, 2014.
59. Nguyen, D.-H.T., S.C. Stapleton, M.T. Yang, S.S. Cha, C.K. Choi, P.A. Galie, et al. Biomimetic model to reconstitute angiogenic sprouting morphogenesis in vitro. Proceedings of the National Academy of Sciences 110, 6712, 2013.
60. Meer, A.D.v.d., V.V. Orlova, P.t. Dijke, A.v.d. Berg, and C.L. Mummery. Three-dimensional co-cultures of human endothelial cells and embryonic stem cell-derived pericytes inside a microfluidic device. Lab on a Chip 13, 3562, 2013.
61. Koh, W., A.N. Stratman, A. Sacharidou, and G.E. Davis, *Chapter 5 In Vitro Three Dimensional Collagen Matrix Models of Endothelial Lumen Formation During Vasculogenesis and Angiogenesis*. Methods in Enzymology. Academic Press. 2008. 83-101.
62. Vickerman, V. and R.D. Kamm. Mechanism of a flow-gated angiogenesis switch: early signaling events at cell-matrix and cell-cell junctions. Integr Biol (Camb) 4, 863, 2012.
63. Sleeboom, J.J.F., H. Eslami Amirabadi, P. Nair, C.M. Sahlgren, and J.M.J. den Toonder. Metastasis in context: modeling the tumor microenvironment with cancer-on-a-chip approaches. Disease Models & Mechanisms 11, 2018.
64. Prabhakarpandian, B., M.-C. Shen, J.B. Nichols, C.J. Garson, I.R. Mills, M.M. Matar, et al. Synthetic Tumor Networks for Screening Drug Delivery Systems. Journal of controlled release : official journal of the Controlled Release Society 201, 49, 2015.
65. Pagano, G., M. Ventre, M. Iannone, F. Greco, P.L. Maffettone, and P.A. Netti. Optimizing design and fabrication of microfluidic devices for cell cultures: An effective approach to control cell microenvironment in three dimensions. Biomicrofluidics 8, 2014.
66. Whiteside, T. The tumor microenvironment and its role in promoting tumor growth. Oncogene 27, 5904, 2008.
67. Ungefroren, H., S. Sebens, D. Seidl, H. Lehnert, and R. Hass. Interaction of tumor cells with the microenvironment. Cell Communication and Signaling : CCS 9, 18, 2011.
68. Senthebane, D.A., A. Rowe, N.E. Thomford, H. Shipanga, D. Munro, M.A.M. Al Mazeedi, et al. The Role of Tumor Microenvironment in Chemoresistance: To Survive, Keep Your Enemies Closer. International Journal of Molecular Sciences 18, 2017.
69. Schaaf, M.B., A.D. Garg, and P. Agostinis. Defining the role of the tumor vasculature in antitumor immunity and immunotherapy. Cell Death & Disease 9, 115, 2018.
70. Reid, S.E., E.J. Kay, L.J. Neilson, A.-T. Henze, J. Serneels, E.J. McGhee, et al. Tumor matrix stiffness promotes metastatic cancer cell interaction with the endothelium. The EMBO Journal, e201694912, 2017.
71. Mendoza, E., R. Burd, P. Wachsberger, and A.P. Dicker, *Normalization of Tumor Vasculature and Improvement of Radiation Response by Antiangiogenic Agents*. Antiangiogenic Agents in Cancer Therapy. Humana Press. 2008. 311-321.
72. Castells, M., B. Thibault, J.-P. Delord, and B. Couderc. Implication of Tumor Microenvironment in Chemoresistance: Tumor-Associated Stromal Cells Protect Tumor Cells from Cell Death. International Journal of Molecular Sciences 13, 9545, 2012.
73. Paszek, M.J., N. Zahir, K.R. Johnson, J.N. Lakins, G.I. Rozenberg, A. Gefen, et al. Tensional homeostasis and the malignant phenotype. Cancer Cell 8, 241, 2005.

74. Privratsky, J.R. and P.J. Newman. PECAM-1: regulator of endothelial junctional integrity. Cell and tissue research 355, 607, 2014.
75. Grainger, S.J. and A.J. Putnam. Assessing the permeability of engineered capillary networks in a 3D culture. PLoS One 6, e22086, 2011.
76. Crosby, C.O. and J. Zoldan. An In Vitro 3D Model and Computational Pipeline to Quantify the Vasculogenic Potential of iPSC-Derived Endothelial Progenitors. JoVE (Journal of Visualized Experiments), e59342, 2019.
77. Kerschnitzki, M., P. Kollmannsberger, M. Burghammer, G.N. Duda, R. Weinkamer, W. Wagermaier, et al. Architecture of the osteocyte network correlates with bone material quality. Journal of Bone and Mineral Research: The Official Journal of the American Society for Bone and Mineral Research 28, 1837, 2013.
78. Kollmannsberger, P., M. Kerschnitzki, F. Repp, W. Wagermaier, R. Weinkamer, and P. Fratzl. The small world of osteocytes: connectomics of the lacuno-canalicular network in bone. New Journal of Physics 19, 073019, 2017.
79. Debeb, B.G., L. Lacerda, S. Anfossi, P. Diagaradjane, K. Chu, A. Bambhroliya, et al. miR-141-Mediated Regulation of Brain Metastasis From Breast Cancer. J Natl Cancer Inst 108, 2016.
80. Colpaert, C.G., P.B. Vermeulen, I. Benoy, A. Soubry, F. van Roy, P. van Beest, et al. Inflammatory breast cancer shows angiogenesis with high endothelial proliferation rate and strong E-cadherin expression. Br J Cancer 88, 718, 2003.
81. Mahooti, S., K. Porter, M.L. Alpaugh, Y. Ye, Y. Xiao, S. Jones, et al. Breast carcinomatous tumoral emboli can result from encircling lymphovasculogenesis rather than lymphovascular invasion. Oncotarget 1, 131, 2010.
82. Azzi, S., J.K. Hebda, and J. Gavard. Vascular permeability and drug delivery in cancers. Front Oncol 3, 211, 2013.
83. Claesson-Welsh, L. Vascular permeability--the essentials. Ups J Med Sci 120, 135, 2015.
84. Hashizume, H., P. Baluk, S. Morikawa, J.W. McLean, G. Thurston, S. Roberge, et al. Openings between defective endothelial cells explain tumor vessel leakiness. Am J Pathol 156, 1363, 2000.
85. Jain, R.K., J.D. Martin, and T. Stylianopoulos. The role of mechanical forces in tumor growth and therapy. Annu Rev Biomed Eng 16, 321, 2014.
86. Shenoy, A.K. and J. Lu. Cancer cells remodel themselves and vasculature to overcome the endothelial barrier. Cancer Lett 380, 534, 2016.
87. Uldry, E., S. Faes, N. Demartines, and O. Dormond. Fine-Tuning Tumor Endothelial Cells to Selectively Kill Cancer. Int J Mol Sci 18, 2017.
88. Lee, H., S. Kim, M. Chung, J.H. Kim, and N.L. Jeon. A bioengineered array of 3D microvessels for vascular permeability assay. Microvasc Res 91, 90, 2014.
89. Tang, Y., F. Soroush, J.B. Sheffield, B. Wang, B. Prabhakarpandian, and M.F. Kiani. A Biomimetic Microfluidic Tumor Microenvironment Platform Mimicking the EPR Effect for Rapid Screening of Drug Delivery Systems. Sci Rep 7, 9359, 2017.
90. Kim, S., H. Lee, M. Chung, and N.L. Jeon. Engineering of functional, perfusable 3D microvascular networks on a chip. Lab Chip 13, 1489, 2013.
91. Jeon, J.S., I.K. Zervantonakis, S. Chung, R.D. Kamm, and J.L. Charest. In vitro model of tumor cell extravasation. PLoS One 8, e56910, 2013.
92. Lee, H., W. Park, H. Ryu, and N.L. Jeon. A microfluidic platform for quantitative analysis of cancer angiogenesis and intravasation. Biomicrofluidics 8, 054102, 2014.
93. Terrell-Hall, T.B., A.G. Ammer, J.I. Griffith, and P.R. Lockman. Permeability across a novel microfluidic blood-tumor barrier model. Fluids Barriers CNS 14, 3, 2017.

94. Zervantonakis, I.K., S.K. Hughes-Alford, J.L. Charest, J.S. Condeelis, F.B. Gertler, and R.D. Kamm. Three-dimensional microfluidic model for tumor cell intravasation and endothelial barrier function. Proc Natl Acad Sci U S A 109, 13515, 2012.
95. Kim, S., W. Kim, S. Lim, and J.S. Jeon. Vasculature-On-A-Chip for In Vitro Disease Models. Bioengineering (Basel) 4, 2017.
96. Brenner, W., P. Langer, F. Oesch, C.J. Edgell, and R.J. Wieser. Tumor cell--endothelium adhesion in an artificial venule. Anal Biochem 225, 213, 1995.
97. Haidari, M., W. Zhang, A. Caivano, Z. Chen, L. Ganjehei, A. Mortazavi, et al. Integrin alpha2beta1 mediates tyrosine phosphorylation of vascular endothelial cadherin induced by invasive breast cancer cells. J Biol Chem 287, 32981, 2012.
98. Haidari, M., W. Zhang, and K. Wakame. Disruption of endothelial adherens junction by invasive breast cancer cells is mediated by reactive oxygen species and is attenuated by AHCC. Life Sci 93, 994, 2013.
99. Kebers, F., J.M. Lewalle, J. Desreux, C. Munaut, L. Devy, J.M. Foidart, et al. Induction of endothelial cell apoptosis by solid tumor cells. Exp Cell Res 240, 197, 1998.
100. Mierke, C.T. Cancer cells regulate biomechanical properties of human microvascular endothelial cells. J Biol Chem 286, 40025, 2011.
101. Peyri, N., M. Berard, F. Fauvel-Lafeve, V. Trochon, B. Arbeille, H. Lu, et al. Breast tumor cells transendothelial migration induces endothelial cell anoikis through extracellular matrix degradation. Anticancer Res 29, 2347, 2009.
102. Zhang, H., C.C. Wong, H. Wei, D.M. Gilkes, P. Korangath, P. Chaturvedi, et al. HIF-1-dependent expression of angiopoietin-like 4 and L1CAM mediates vascular metastasis of hypoxic breast cancer cells to the lungs. Oncogene 31, 1757, 2012.
103. Vermeulen, P.B., K.L. van Golen, and L.Y. Dirix. Angiogenesis, lymphangiogenesis, growth pattern, and tumor emboli in inflammatory breast cancer: a review of the current knowledge. Cancer 116, 2748, 2010.
104. Shirakawa, K., M. Shibuya, Y. Heike, S. Takashima, I. Watanabe, F. Konishi, et al. Tumor-infiltrating endothelial cells and endothelial precursor cells in inflammatory breast cancer. Int J Cancer 99, 344, 2002.
105. van Golen, K.L., Z.F. Wu, X.T. Qiao, L. Bao, and S.D. Merajver. RhoC GTPase overexpression modulates induction of angiogenic factors in breast cells. Neoplasia 2, 418, 2000.
106. Guzman, A., M.J. Ziperstein, and L.J. Kaufman. The effect of fibrillar matrix architecture on tumor cell invasion of physically challenging environments. Biomaterials 35, 6954, 2014.
107. Holle, A.W., J.L. Young, and J.P. Spatz. In vitro cancer cell-ECM interactions inform in vivo cancer treatment. Adv Drug Deliv Rev 97, 270, 2016.
108. Lautscham, L.A., C. Kammerer, J.R. Lange, T. Kolb, C. Mark, A. Schilling, et al. Migration in Confined 3D Environments Is Determined by a Combination of Adhesiveness, Nuclear Volume, Contractility, and Cell Stiffness. Biophys J 109, 900, 2015.
109. Sabeh, F., R. Shimizu-Hirota, and S.J. Weiss. Protease-dependent versus -independent cancer cell invasion programs: three-dimensional amoeboid movement revisited. J Cell Biol 185, 11, 2009.
110. Seo, B.R., P. DelNero, and C. Fischbach. In vitro models of tumor vessels and matrix: engineering approaches to investigate transport limitations and drug delivery in cancer. Adv Drug Deliv Rev 69-70, 205, 2014.
111. Wolf, K. and P. Friedl. Extracellular matrix determinants of proteolytic and non-proteolytic cell migration. Trends Cell Biol 21, 736, 2011.
112. Al-Raawi, D., H. Abu-El-Zahab, M. El-Shinawi, and M.M. Mohamed. Membrane type-1 matrix metalloproteinase (MT1-MMP) correlates with the expression and activation of matrix metalloproteinase-2 (MMP-2) in inflammatory breast cancer. Int J Clin Exp Med 4, 265, 2011.

113. Lang, N.R., K. Skodzek, S. Hurst, A. Mainka, J. Steinwachs, J. Schneider, et al. Biphasic response of cell invasion to matrix stiffness in three-dimensional biopolymer networks. Acta Biomater 13, 61, 2015.
114. Sabeh, F., I. Ota, K. Holmbeck, H. Birkedal-Hansen, P. Soloway, M. Balbin, et al. Tumor cell traffic through the extracellular matrix is controlled by the membrane-anchored collagenase MT1-MMP. J Cell Biol 167, 769, 2004.
115. Wolf, K., M. Te Lindert, M. Krause, S. Alexander, J. Te Riet, A.L. Willis, et al. Physical limits of cell migration: control by ECM space and nuclear deformation and tuning by proteolysis and traction force. J Cell Biol 201, 1069, 2013.
116. Wolf, K., Y.I. Wu, Y. Liu, J. Geiger, E. Tam, C. Overall, et al. Multi-step pericellular proteolysis controls the transition from individual to collective cancer cell invasion. Nat Cell Biol 9, 893, 2007.
117. Rizwan, A., M. Cheng, Z.M. Bhujwalla, B. Krishnamachary, L. Jiang, and K. Glunde. Breast cancer cell adhesome and degradome interact to drive metastasis. npj Breast Cancer 1, 15017, 2015.
118. Kleer, C.G., K.L.v. Golen, T. Braun, and S.D. Merajver. Persistent E-Cadherin Expression in Inflammatory Breast Cancer. Modern Pathology 14, 458, 2001.
119. Sugino, T., T. Kusakabe, N. Hoshi, T. Yamaguchi, T. Kawaguchi, S. Goodison, et al. An Invasion-Independent Pathway of Blood-Borne Metastasis. The American Journal of Pathology 160, 1973, 2002.
120. Carmeliet, P. VEGF as a key mediator of angiogenesis in cancer. Oncology 69 Suppl 3, 4, 2005.
121. Hoeben, A., B. Landuyt, M.S. Highley, H. Wildiers, A.T. Van Oosterom, and E.A. De Bruijn. Vascular endothelial growth factor and angiogenesis. Pharmacol Rev 56, 549, 2004.
122. Wolfe, A.R., N.J. Trenton, B.G. Debeb, R. Larson, B. Ruffell, K. Chu, et al. Mesenchymal stem cells and macrophages interact through IL-6 to promote inflammatory breast cancer in pre-clinical models. Oncotarget 7, 82482, 2016.

**Figure Legend**

**Fig. 1** Development of a confluent endothelium in the vessel using the graded flow protocol in various co-culture conditions. a) Progression of endothelium alignment and confluence throughout the flow protocol for the TIME only platform. The 0 hour time point, imaged after channel formation, initiated with TIME cells in a rounded morphology. The subsequent 48 hour of flow promoted TIME cell spreading and proliferation followed by alignment of the TIME cells in the direction of flow. The resulting confluent endothelium at 78 hours serves to function as a barrier for transendothelial flow; scale bar: 200 µm. b) The resulting *in vitro* vascularized breast tumor platforms consisting of monoculture of TIME cell seeded lumen (red) or co-culture of GFP labeled (green) MDA-IBC3, SUM149, and MDA-MB-231 tumor cells around a TIME cell seeded lumen (red); scale bar: 500 µm.

**Fig. 2** Characterization of the *in vitro* vascularized microfluidic platforms. (a) Endothelial morphology and adhesion observed through PECAM-1 and DAPI (top row, scale bar: 100 µm), and F-actin and DAPI (middle row, scale bar: 200 µm) immunofluorescent staining, and SEM analysis of the endothelium (bottom, scale bar: 10 µm). PECAM-1 (green) staining revealed difference in endothelial cell-cell junctions between neighboring TIME cells, F-actin (red) staining and SEM images revealed morphological difference in the endothelium. White arrows denote gaps between the neighboring cells and the boxed areas in the F-actin images show early signs of angiogenic sprouting. (b) Quantification of endothelium coverage of the vessel lumen from F-actin stained images revealed a decrease in coverage in the MDA-MB-231 and SUM149 platforms; $*p<0.05$, $** p< 0.01$. (c) Measured effective permeability of 70 kDA green fluorescent dextran perfusion through the platforms showed a significant decrease in vessel permeability in the MDA-MB-231 and SUM149 platforms; $*p<0.05$. (d) VEGF and bFGF expression measured by ELISA showed significantly higher VEGF expression in the IBC platforms while bFGF was higher in the non-IBC and acellular control platforms; ■, ▲ denote significance ($p<0.05$) compared to acellular TIME control and MDA-MB-231 platforms respectively.

**Fig. 3** Collagen porosity measured with SEM. a) SEM images of tumor (scale bar: 2 µm), and TIME (scale bar: 20 µm) cells morphologies (top panels), and collagen matrix organization (bottom panels, scale bar: 1 µm). b) Collagen matrix porosity measurements calculated from SEM images of the ECM, $*p<0.05$, $**p<0.01$

**Fig. 4** Vascular sprouting dynamically observed over a three week period in the MDA-IBC3/TIME *in vitro* vascularized tumor platforms. (a) Longitudinal cross section images of the vessel show vessel sprouting, branching, as well formation of tumor emboli pointed out by white arrows (top panels), and front view of the vessels (bottom panels). (b) K-S analysis of vessel sprouting revealed a significant increase in sprouting at later time points compared to Day 0. (c) F-actin (red) staining of GFP labeled MDA-IBC3 cells (green) showed formation and growth of tumor emboli. (d) Lumen formation followed over time in one of the vessel sprouts. (e) Vascular nesting phenomenon of IBC tumors in *in vivo* patient derived histological samples demonstrated by CD31 staining of vascular vessel (brown) surrounding IBC tumor emboli (blue). (f) *In vitro* recreation of vascular nesting of IBC tumors as shown by the encircling of MDA-IBC3 tumor cells (green) by mKate labeled sprouts (red).

**Fig. 5** Confirmation of lumen formation in the vessel sprouts on Day 14 (a) and Day 21 (b). Platforms were injected with a solution of 1µm green fluorescent microspheres and presence of green microspheres away from the main vessel was indicative of lumen formation in the sprouts.

**Fig. 6** Analysis of the vascular network. (a) A 45 µm region of interest at the center of the vascular vessel was used to quantify the number of sprouts and the network length. (b) Representative images of the vascular network in the region of interest derived by applying the algorithm developed by Kollmannsberger et al. and Crosby et al. [76-78] used for determining (c) fold change comparisons of total network length and number of sprouts normalized to the Day 0 values. Results show a steady increase in both number of sprouts and total vascular network length at each subsequent time point. (d) A schematic cross-section of the vessel with vascular sprouting used for calculation of number of sprouts with cross sectional areas ranging from 100 to 1000 µm$^2$ determined at 100, 200, 300 and 400 µm away from the vascular vessel. (e) Number of sprouts present at 100, 200, 300, and 400 µm away from the vessel as well as the cross sectional areas of those sprouts were determined. Over time, the new sprouts increase in both cross sectional area, an indication of lumen formation, and length. (f) Measurement of the diameters of the *in vitro* sprouts with lumen capable of particle perfusion (vessels with green signal from microspheres overlaying red signal from endothelial sprouts) taken on Day 12. Sprout area range from 15-42 µm (scale bar indicates 100 µm) and correlate with the increase in larger area sprouts found at later time points in Fig. 6e.

**Fig. 7** Cytokine analysis of angiogenic associated factors measured over a three week period. ANG2, VEGF-A, PDGF-bb, IL-8, IL-6, and MMP2 showed a significant increase in expression on Day 21 compared to earlier timepoints while bFGF and EGF both peaked on Day 7, *$p<0.05$.

**Supplementary Information**

*S1. Sprouting Quantification Methodology*
Analysis for Vessel Quantification: We have developed an in-house algorithm for quantifying angiogenesis in the platform. We map the experimental vessel that is a cylinder with a near elliptical cross section onto a true cylinder. This is reasonable since a ~ 0.95b, where a is the semi-major axis and b is the semi-minor axis of the elliptical cross section. We then map the elliptical cross section onto a circular cross section and assume this preserves area since a is nearly equal to b. Since the true cylindrical vessel has radius R and we are interested in quantifying vascular sprouting at distances of 50, 100, 150, etc. from the vessel, we create cylindrical shells that are 50, 100, 150, etc. microns from the experimental vessel. We construct this surface using intensity values of fluorescence and unroll the cylindrical shells into rectangular surfaces. We use an intensity threshold on these surfaces and then quantify vessel sprouting by taking area thresholds of varying size that correspond to the cross-sectional area of the vessel going through these constructed surfaces at varying distances from the parent vessel.

**Supplementary Fig. A**: Calcein staining (red) of live GFP tumor cells (green) in the platforms without TIME cells 12 hours after initial seeding to visualize cell numbers; scale bar 300μm. All cells were seeded at an initial density of 1 million cells/ml. Difference in inherent GFP expression are shown as the IBC cells lines MDA-IBC3 and SUM149 express a strong GFP signal while the GFP expression of MDA-MB-231 cells is much weaker.